# Paraelectric in a Strong High-Frequency Field

A. I. Sokolov

*St. Petersburg Electrotechnical University, St. Petersburg, 197376 Russia*
E-mail: ais2002@mail.ru

**Abstract**—A change in the effective permittivity of a ferroelectric film in the paraelectric phase under the action of a strong high-frequency field (nonequilibrium soft mode heating) is considered. It is shown that this effect must be most clearly pronounced far from the resonance ($\omega_0 \ll \omega_{sm}$), rather than for the external field frequency $\omega_0$ close to the soft mode frequency $\omega_{sm}$. The effective permittivity as a function of the high-frequency field amplitude is determined using a phenomenological approach and within the framework of a microscopic theory based on a simple model of a displacement-type ferroelectric.

The first attempts to use ferroelectrics in the paraelectric phase as nonlinear media at high frequencies were undertaken in the 1960s [1, 2]. It was established that the main obstacle to such applications is related to high dielectric losses. In recent decades, these losses were reduced to a quite acceptable level, while the controllability (i.e., the dependence of the permittivity on the external electric field) was retained. This circumstance led to the revival of interest in paraelectrics. However, the development of thin-film nonlinear elements based on these compounds showed that their response to constant and alternating external fields can be substantially different [3]. In particular, the switching (changing capacitance) in ferroelectric capacitors by an external pulsed control field may proceed with anomalously large relaxation times reaching seconds and minutes at room temperature. This behavior makes the technical application of such nonlinear elements problematic.

The aforementioned anomaly is most likely related to the slow relaxation of the charge appearing at the defects (traps) in the near-electrode regions under the action of a control voltage. The discharge of these regions can be accelerated—and, hence, the response speed of nonlinear elements increased—by irradiating a capacitor with light possessing a photon energy that is greater than the bandgap width [4]. Another method of eliminating this difficulty consists in using a high-frequency alternating electric field, rather than dc voltage (video pulses) as control signals. If the frequency of this field is close to that of a natural soft mode ($10^{11}$–$10^{12}$ Hz) and the amplitude is sufficiently large, the phonon subsystem of the material will pass to a nonequilibrium state corresponding to soft-mode "overheating" and, hence, the permittivity $\varepsilon$ will exhibit a rapid change [5].

The present investigation was aimed at establishing frequencies of such an external signal, which would provide the most effective control over a nonlinear element at a minimum heating of the material by the high-frequency field. It should be noted that the direct (quasi-equilibrium) heating of the element by an external field also leads to a change in the permittivity and can be considered as a means of controlling the capacitance. However, the characteristic times of such thermal processes even in thin-film samples are several orders of magnitude greater than the typical lattice relaxation times ($\sim 10^{-11}$ s) for paraelectrics in the vicinity of the phase transition temperature $T_c$, so that this method of control does not allow the potentially high response speed to be realized.

Thus, the question is whether it is necessary to act upon the crystal subsystem responsible for the ferroelectric phase transition (soft mode) at the resonance frequency? The answer would certainly be positive, if the soft mode exhibited pronounced resonance behavior. Unfortunately, even in the seldom cases where the ferroelectric mode is not overdamped, its resonance character is rather weakly manifested. For example, at temperatures close to $T_c$, an increase in the real part $\varepsilon'$ of the permittivity of high-quality single crystals entering the range of dispersion does not usually exceed several dozen percents. In ferroelectrics of average quality (polycrystalline, ceramic, etc.), the frequency dependence $\varepsilon'(\omega)$ is typically described by almost monotonically decreasing curve. This behavior implies that, by applying a control field with a frequency of $\omega_0 \approx \omega_{sm}$ to a ferroelectric nonlinear element, we will not obtain any significant advantage as compared to the control at any other frequency $\omega_0 \ll \omega_{sm}$.

On the other hand, an increase in the frequency of a control field in the (0, $\omega_{sm}$) interval will unavoidably lead to enhanced heating of the sample because both the imaginary part $\varepsilon''$ of the permittivity and the dielectric loss tangent (tan$\delta$) grow with the frequency. Thus, the choice of control frequency $\omega_0$ close to $\omega_{sm}$ can hardly be considered optimum, since the direct heating gives the maximum contribution to a change in the permittivity and masks the role of a dielectric nonlinearity.

Evidently, the situation can be improved by decreasing $\omega_0$. However, this decrease cannot be arbitrary because the working frequency $\omega$ (at which the permittivity $\varepsilon$ is measured) must be significantly smaller than the control frequency. Only this circumstance allows us to believe that a paraelectric possesses certain permittivity dependent on the amplitude $E_0$ of the control field. Thus, in what follows we will assume that $\omega \ll \omega_0 \ll \omega_{sm}$. Let us obtain in this case an expression for the effective permittivity $\varepsilon_{eff}(\omega)$ as a function of the field amplitude $E_0$. This problem is quasi-static and the dissipative effects can be ignored. Under these conditions, we can assume that $\varepsilon_{eff}(\omega) = \varepsilon_{eff}(0)$ and calculate the static effective permittivity $\varepsilon_{eff} = \varepsilon_{eff}(0)$ using the Landau expansion of the free energy in powers of the polarization $P$:

$$F = \frac{1}{2} a P^2 + \frac{1}{4} b P^4. \qquad (1)$$

where $a = \alpha(T - T_C)$, $b > 0$, and both $\alpha$ and $b$ are assumed to be independent of the temperature. In the paraelectric phase, $a > 0$ and one can readily check that $1/a = \chi \approx \varepsilon$. If the material is exposed to an alternating external field, the polarization can be expressed as $P + P_0 \cos(\omega_0 t)$. Substituting this expression into relation (1) and averaging over the field period $2\pi/\omega_0$, we eventually obtain the following formula:

$$\varepsilon_{eff} = \varepsilon \left[ 1 + (3/2) b \varepsilon P_0^2 \right]^{-1}. \qquad (2)$$

Expressing $P_0$ through the control field amplitude, we obtain the following formula in the lowest approximation with respect to $E_0$:

$$\varepsilon_{eff} = \varepsilon - (3/2) b \varepsilon^4 E_0^2. \qquad (3)$$

A change in the permittivity of a ferroelectric film in the paraelectric phase under the action of a strong high-frequency field can also be determined within the framework of a microscopic theory. Let us consider a displacement-type ferroelectric, the thermodynamics and dynamics of which can be described using an anharmonic model with an imaginary seeding frequency of the soft mode $\omega_{sm}^2(0, 0) = -\alpha^2 = -T_c/C$. The Hamiltonian of this model is as follows:

$$H = (2\varepsilon_0)^{-1} \left[ \sum_{\mathbf{q}} \left( -\alpha^2 + s^2 q^2 \right) \varphi(\mathbf{q}) \varphi(-\mathbf{q}) \right. $$
$$\left. + (\beta/2) \sum_{\mathbf{qq'q''}} \varphi(\mathbf{q}) \varphi(\mathbf{q'}) \varphi(\mathbf{q''}) \varphi(-\mathbf{q} - \mathbf{q'} - \mathbf{q''}) \right]. \quad (4)$$

where $C$ is the Curie constant, $\varepsilon_0$ is the permittivity of vacuum, $\beta$ is the anharmonicity constant, and $\varphi(\mathbf{q})$ is the Fourier component of the polarization fluctuation field. The summation over wave vectors $\mathbf{q}$, $\mathbf{q'}$, $\mathbf{q''}$ in Eq. (4) is performed within the first Brillouin zone. Assuming for simplicity the frequency $\omega_{sm}(\mathbf{q}, T)$ to be dimensionless and using the quasi-anharmonic approximation (or self-consistent phonon approximation) with averaging over thermal fluctuations, we readily obtain the following relation [6, 7]:

$$\varpi_{sm}^2(\mathbf{q}, T) = -\alpha^2 + s^2 q^2 + 3\beta \sum_{\mathbf{q'}} \langle \varphi(\mathbf{q'}) \varphi(-\mathbf{q'}) \rangle =$$
$$\varpi_{sm}^2(0, T) + s^2 q^2 \qquad (5)$$

where $\langle \ldots \rangle$ denotes the thermodynamic average. Since the pair correlator at not too low temperatures can be expressed as $\langle \varphi(\mathbf{q'}) \varphi(-\mathbf{q'}) \rangle = 2\varepsilon_0 k_B T / \varpi_{sm}^2(\mathbf{q'}, T)$, Eq. (5) describes the soft mode frequency as a function of the temperature:

$$\varpi_{sm}^2(0, T) = -\alpha^2 + 6\beta \varepsilon_0 k_B T \sum_{\mathbf{q'}} \varpi_{sm}^{-2}(\mathbf{q'}, T), \quad (6)$$

where $k_B$ is the Boltzmann constant. Replacing the sum in Eq. (6) by an integral with the density of states $(2\pi)^{-3}$ and solving this equation in the limiting case of $\omega_{sm}^2(0, T) \ll s^2 q_D^2$, where $q_D$ is the cutoff momentum, we eventually obtain

$$\varpi_{sm}^2(0, T) = (T - T_c)/C, \; C = \pi^2 s^2 (3\beta q_D k_B \varepsilon_0)^{-1}. \quad (7)$$

Note that, at $\mathbf{q} = 0$ and $\omega \ll \omega_{sm}$, the frequency $\omega_{sm}(0, T)$ determines the permittivity of a paraelectric as

$$\varepsilon = \varepsilon_0 \varpi_{sm}^{-2}(0, T). \qquad (8)$$

According to Eq. (5), the frequency $\omega_{sm}(0, T)$ depends on the intensity of polarization fluctuations. If the sample is exposed to a high-frequency external field, a non-equilibrium contribution related to this field is added to the thermodynamic average $\sum_{\mathbf{q'}} \langle \varphi(\mathbf{q'}) \varphi(-\mathbf{q'}) \rangle$ in

Eq. (5). For the simplicity, let us consider a spatially homogeneous alternating field. Then, the nonequilibrium polarization component can be expressed as follows:

$$\mathrm{Re}\,\varphi_{ne}(\mathbf{q}) = \varphi_{ne}(0)\delta_{0,\mathbf{q}} = P_0 \delta_{0,\mathbf{q}}, \quad (9)$$

where $\delta_{0,q}$ is the Kronecker delta. A contribution of this component to the right-hand part of Eq. (5) is $3\beta P_0^2/2$. As a result, we obtain the following expressions for the frequency $\omega_{sm}$ of a soft mode heated by the high-frequency and the corresponding effective permittivity

$$\varpi_{sm}^2(0,T,P_0) = (T - T_c)/C + (3/2)\beta P_0^2, \quad (10)$$

$$\begin{aligned}\varepsilon_{eff}(P_0) &= \varepsilon_0 \varpi_{sm}^2(0,T,P_0) \\ &= \varepsilon[1 + (3/2)\beta\varepsilon\varepsilon_0^{-1} P_0^2]^{-1}\end{aligned}, \quad (11)$$

As can readily be seen, expression (11) is equivalent to formula (2) obtained above using a phenomenological approach.

**Acknowledgments**. The author is grateful to A.M. Prudan for attracting attention to the problem under consideration.

This study was supported in part by the Ministry of Education and Science of the Russian Federation (project no. RNP.2.1.2.7083) and the Russian Foundation for Basic Research (project no. 07-02-00345).